\documentclass{article}

\usepackage[preprint]{neurips_2021}

\usepackage[utf8]{inputenc} 
\usepackage[T1]{fontenc}    
\usepackage{hyperref}       
\usepackage{url}            
\usepackage{booktabs}       
\usepackage{amsfonts}       
\usepackage{nicefrac}       
\usepackage{microtype}      
\usepackage{xcolor}         
\usepackage{wrapfig}
\usepackage{bm}
\usepackage{amsmath}
\usepackage{amssymb}
\usepackage{amsthm}
\usepackage{amsmath}
\usepackage[inline]{enumitem}
\usepackage{algorithm}
\usepackage[noend]{algpseudocode}
\usepackage{graphicx}
\usepackage{diagbox}
\usepackage{multirow}
\usepackage{apptools}
\usepackage{diagbox}
\usepackage{makecell}
\usepackage{mathtools}
\usepackage{caption}
\usepackage{subcaption}

\title{Towards a Fairer Digital Marketing Model}

\author{
Leo Ardon \\
Imperial College London\\
\And
Dario Morelli \\
Imperial College London\\
\And
Francesco Villani \\
Imperial College London\\
\And
David Wheatley\\
Imperial College London\\
}

\begin{document}

\maketitle

\begin{abstract}
    Surfing on the internet boom, the digital marketing industry has seen an exponential growth in the recent years and is often at the origin of the financial success of the biggest tech firms. In this paper we study the current landscape of this industry and comment on the monopoly that Google has managed to gain over the years through technical innovations and intelligent acquisitions. We then propose potential avenues to explore in an effort to help moving the digital marketing industry towards a fairer model.
\end{abstract}

\section{Introduction}

\subsection{Google and Alphabet}

Google was officially launched in 1998 by Larry Page and Sergey Brin, marking the beginning of its rise to becoming the 5th most valuable company in the world with a market capitalisation of 1.438 trillion dollars as of April 2021 \citep{YChart_2021}.

While the primary objective of almost all for-profit companies is to deliver shareholder value through growth, both via strategic acquisition and organically, questions have been raised regarding the ethics of Google’s monumental success. In particular, the near-monopolistic position Google now holds in the marketing world has resulted in multiple antitrust cases being presented against them and forms the discussion of this report.

This report has been split into 5 main sections.
\begin{enumerate}
    \item In this section we provide some background to the case, legislation and state our position with regards to the antitrust case.
    \item We then explore the tools that Google have developed providing them with the data that support their unique ability to profile and target individuals.
    \item Here we provide some context to the digital marketing industry, again focusing on how Google have obtained their dominant position.
    \item We then investigate the wider implications Google’s market dominance has had on other industries.
    \item Finally we propose technological and legislative solutions that we believe could help deliver a more fair and open market.
\end{enumerate}

\subsection{Background to the antitrust case}

The Investopedia Guide to Antitrust laws \citep{Investopedia_Antitrust} provides the following definition that we will use to frame our discussion: “Antitrust laws are regulations that encourage competition by limiting the market power of any particular firm”. They go on to explain that “due to the complexity of deciding what rules will restrict competition, antitrust law has become a distinct legal specialisation.”

Looking more specifically at the Google case, we refer to an article in the Independent newspaper \citep{Independent_2021} that states that “Google has long defended itself against monopoly charges by stressing that its products are free and that no one has to use them.” The article goes on to state that “the users of Google search engines, browsers and other services are (however) not the real customer, but the product ... advertisers are its real customers”. It is Google's ability to leverage this user data, allied with its dominance of the digital marketing industry, that has led to the three antitrust cases currently being pursued by the Department of Justice in the US, with similar cases being filed across Europe. These have been defined in the technology magazine Wired \citep{Wired_2021}:
\begin{enumerate}
    \item The first claims that “Google has used anti-competitive tactics to protect its monopoly over general search and prevent rival search engines from getting a foothold”.
    \item The second argues that “Google has made changes over the years to how search results appear to keep more traffic flowing to Google's properties rather than vertical search, like Yelp or Kayak.”
    \item The third, and probably the most serious, claims that “Google exploits its control over the advertising pipeline to impose unfair conditions on advertisers and publishers, discriminate against rival ad tech firms and generate a much larger share of online ad spending profits than it
would earn if more firms were competing in the market.”
\end{enumerate}

\subsection{Data legislation, ownership and controls}

In April 2016, the European Union passed the General Data Protection Regulation (GDPR) to establish how companies and organisations should process data related to EU residents. Critically, if a company operates outside the EU but serves users within it, GDPR is still applicable: this makes the regulation de facto one with global reach.

At its heart, GDPR aimed to strike a balance between easing the international transfer of data between the Member States and protecting the users to whom the data belongs. For the first time, it set the key principle that users own their data and should therefore explicitly give consent to allow data collection and processing.

Cookies are a core element of online personalised advertising, and article 30 of the GDPR text recognises that "Natural persons may be associated with online identifiers provided by their devices, applications, tools and protocols, such as internet protocol addresses, cookie identifiers or other identifiers such as radio frequency identification tags." \citep{Reg_2016}.

To meet compliance requirements, websites must now present users with cookie banners once a page is loaded, in line with the European Data Protection Board guidelines released in May 2020 \citep{EDPB_2020}: websites' visitors can then accept all, some or none of the first- and third-party cookies.

Against this regulatory background, in September 2020 CNIL, the French Data Protection Authority, fined Google €50 million (\$57 million) for breaching GDPR rules \citep{CNIL_2019}. Specifically, Google was accused of failing to: (i) inform users that their consent was mainly for the purpose of ad personalisation; (ii) clarify which of its internal products, e.g. Google Maps, YouTube, etc., would benefit from the aggregation and processing of the data.

While European watchdogs can undertake initiatives like CNIL's due to GDPR, the regulatory framework in the rest of the world is much more fragmented. The United States do not offer an equivalent of GDPR at the Federal level, although individual states might have approved local data protection laws. Other large economies such as Australia, Japan and Canada have proposed rules in the same spirit as GDPR, but their levels of enforcement vary greatly \citep{Comforte_2021}.

In the current landscape, it can be said that Google's presence is ubiquitous, while data protection for its customers is not, which leaves them exposed to arbitrary policies.

\subsection{Summary of our position}

Google has obtained an unparalleled understanding of user activity through its search, maps, browser or authentication platforms. This data has enabled them to target advertisements more effectively than anyone else and subsequently extract greater value. In this report, we start by assessing the tools Google have developed, and provided to users for free, in exchange for the privileged access to data that powers their marketing machine.

As with many software platforms, the value a user derives from a service increases as the number of active users grows: this is known as a network effect. These effects are prevalent in digital marketing: profiling users across their complete online activity increases the value of an advert, as it can be more targeted and ultimately maximise conversion rates. The question is whether Google’s growth has resulted from the natural tendency of the ad market to monopolise due to network effects or whether Google has unlawfully stifled competition to obtain its market position.

In this report we suggest that, while Google’s actions have not been illegal, the existence of a single company dominating the complete advertising process, and ultimately controlling the monetisation of the internet, is highly undesirable and open to abuse. Specific actions undertaken by Google have aggressively sought to maintain this position. The rapidly evolving state of the industry and the lack of a robust regulatory body have failed to impose the necessary controls to maintain a fair and open market. \cite{srinivasan2020google} uses an analogy of the financial markets stating that “a stock exchange would not be allowed to operate a division involved in trading. It is essential that brokers cannot use information about their customers trading activities for their own financial gain”.

We, therefore, propose that equivalent regulations, to those imposed on the financial markets, are required to ensure a less skewed distribution of the marketing revenues throughout the value chain and stimulate wider innovation across the industry. We also explore the opportunities that emerging technologies, such as blockchain, provide in delivering a more transparent, decentralised and fair model. However, before we begin developing potential solutions, we must first analyse the technologies and markets in which Google has been so successful in expanding its market position.

\section{Data is power}

Collecting, processing and leveraging customers' data is at the heart of Google's success, allowing them to connect users to web content and target adverts better than anyone else. In the following sections, we provide an overview of the primary products which have enabled, and benefited from, this privileged access to data and how an unfair advantage has emerged at the expense of its competitors.

\subsection{Google Search}

Google’s original product was a search engine that revolutionised the way people navigated the internet. Using a new algorithm, called PageRank, and the concept of using back-links to compute the relevance of a page, Google quickly became the leader in the market, helping millions of users navigate the world wide web quickly and efficiently. Google searches can be decomposed into two main capabilities: the indexing of the web and the relevance of the results provided to the user.

Crawling the immensity of the web requires resources that only a few can afford. Google has both the technical expertise and resources to crawl billions of pages and compute the secret score that will help prioritise which pages are displayed is response to a search request. Many other players have tried to compete with Google, developing their own crawling infrastructure and proprietary ranking algorithms, but the cost required to maintain such a tech stack can increasingly only be afforded by technology giants such as Google and Microsoft. To illustrate this point, DuckDuckGo, a search engine focused on protecting user privacy, was forced to stop crawling the web and chose to leverage the Microsoft infrastructure a couple of years after launching due to the escalating cost. Also, the amount of data that Google has collected throughout the years has given them a head-start that none of its competitors can catch up on and is further reinforced by the network effects discussed earlier.

Thanks to the accuracy of results being returned, Google search engine has rapidly become the go-to website, displacing both traditional and specialised search verticals to find anything on the web. The term "Googling" has entered common parlance as a synonym for looking up information on the internet resulting from Google delivering 92\% \citep{StatCounter_2021_SE} of all the searches performed online!

Given its reach, Google Search is able to attract premium ad space prices and represents 63\% of Google's overall revenues \citep{SEC_2019} . Despite Page and Brin's initial position against advertising, the number of ads shown on the Google Search web page has steadily grown, initially limited to text only ads, but more recently also allowing images to also be displayed.

The success of the iPhone and the rapid growth of the mobile market gave an opportunity to Apple to threaten Google's monopolistic position. Google quickly mitigated the risk this posed to their business by signing a lucrative contract with Apple to establish Google as the default search engine on all iOS platforms, safeguarding Google's search engine monopoly. Although the amount paid by Google to Apple keeps increasing every year, both companies benefit enormously.

Google’s original innovative search solution revolutionised the web and has enabled them to remain ahead of the competition ever since. This technology now provides a unique insight into user activity and a lucrative source of revenue through the adverts displayed, which continues to offset the financing required by some of its other, less lucrative ventures.

\subsection{Google Maps}

The earliest prototype of Google Maps was initially developed by the company Where 2 Technologies in 2004. Google Maps was subsequently launched as a standalone application in 2005 following the acquisition of Where 2 Technologies, along with a few other geospatial technology companies. The advent of Google Maps has been seismic: currently, it has over 1 billion active monthly users and together with Google Street View, a service that captures terrestrial images, it has reached near-global coverage. It has allowed people to navigate to new locations more easily around the world and facilitated the deskilling of a workforce as inexperienced drivers, armed with a rented car and a smartphone, are able to compete with traditional taxi firms through a platform such a Uber.

Like many other Google products, Google Maps can be accessed online for free, but the value it generates is in capturing hyper-localised geolocation data and the ability to publish relevant business listings. When users share their location with the service, Google Maps provides an opportunity to display business details within a defined radius selecting only those most relevant based on previously collected data such as saved preferences, browsing history and behavioural patterns. If a shopper has run a Google search on shoes earlier in the morning and finds themselves in a busy street in New York on a Saturday afternoon, Google Maps can use this data to influence which shops it subsequently displays to the user.

Locally targeted advertising is a simple yet powerful strategy to monetise both users' data and businesses' marketing spend. Google further refined this idea by releasing the "Promoted Pin" feature in 2016: this allowed companies to fill in a free Google My Business (GMB) account and then gain additional online presence through ad spending. Advertising businesses pay on a cost-per-click basis and are encouraged to optimise their SEO strategies to increase their chances of standing out in local searches. The combination of individuals' data, geospatial searches and advertising generates a closed loop where Google has an apparent anti-competitive advantage.

Alternative solutions such as OpenStreetMap address the issue of privacy and focus mainly on providing the best map usability experience but, as with browsing, their market penetration is too small to challenge Google. In contrast Waze, a community-based alternative to Google maps centred around real time sharing of information, in 2011 had amassed a user base of 50 million. Favoured by millennials, Waze threatened Google's dominance; however, this was addressed through a \$1.3 billion acquisition, once again securing Google's position as the undisputed leader.

\subsection{Google Chrome}

Google first released its Chrome browser in 2008, at a time when Microsoft's Internet Explorer (IE) had a dominant 65\% worldwide market share. Within ten years, Chrome surpassed IE and has since had a stable market share between 60\% and 70\%, with only Apple's Safari as a realistic contender with just 20\% \citep{StatCounter_2021_Browser}. Just one year after release, Chrome's userbase grew from 0 to 30 million users, primarily due to an enhanced user experience, faster browsing and the availability of customisable add-ons and extensions.

Fundamental to providing an intuitive and consistent experience of web sites is the use of cookies. Cookies are simple text files stored on a user's machine when a website is accessed, enabling information to persist between user sessions. Initially these cookies were designed to save preferences, store status, or customise the navigation experience. However, more recently 3rd party cookies have been used by large advertising agencies to profile individual user surfing habits across multiple sites and enable the delivery of targeted advertising.

While most browsers use cookie technology in a similar way, Google has a privileged position of being able to combine the information obtained from the customers and use it directly to fund its advertising operations, serving three roles simultaneously as an ad exchange provider, as an intermediary for advertisers and as a publisher to its own search engine.

Due to the growth of mobile devices and the associated bandwidth limitations, along with a growing awareness of privacy invasion, ad-blockers began to be more widely adopted, preventing unwanted advertising being displayed on user devices. Since most ads served via Google are charged based on the number of clicks, preventing Chrome from showing the ads could have dramatic consequences. According to a recent report from anti-AdBlock tech firm PageFair, Google lost \$6.6 billion in global revenue to ad blockers in 2014 \citep{BusinessInsider_2015}, accounting for 10\% of Google's total revenues in that year.

The company's strategy changed and in 2020 when it announced that third-party cookies would be slowly phased out in favour of a more privacy friendly approach, which would protect users' identities while preserving the livelihood of the advertising business relying on cookies.

In January 2021, Google released a blog post entitled "Building a privacy-first future for web advertising" \citep{Google_2021}. Therein, it introduced the concept of Federated Learning of Cohorts (FLoC), whereby users are clustered based on their interests instead of being individually identifiable. According to this new protocol, advertisers could still retain most of the conversion they would obtain from cookies, and users could keep their privacy. And while this new practice appears fairer, the applied clustering still relies on users' interests, which Google is uniquely positioned to collect through the access it maintained, via Chrome, to user browsing histories and preferences.

Other browsers such as Firefox and Safari opted for a complete block of third-party cookies in 2019 and 2020, respectively. Another browser, called Brave, approached the problem differently: it put page loading speed and users' privacy at the centre of its mission and built a product around it. Brave blocks all ads and trackers linked to web pages by default, which in turn drastically improves the page loading speed and helps customers protect their data. Using a machine-learning algorithm operating on the users' machines, selected ads are then presented based on the pages visited and search queries performed. Brave claim this approach can outperform third-party tracking in terms of click-through rates, while also ensuring user privacy is not compromised. Brave was first introduced in 2016, and Google's proposed FLoC's model from 2020 appears to have, at least in part, taken inspiration from it. What sets Brave apart is its innovative approach for capturing and distributing value between advertiser, user and publisher. We will discuss this topic in more depth within our proposal, in section \ref{sec:BAT}.

\subsection{Google Authentication}

The suite of Google products discussed above (Search, Maps and Chrome) do not require any form of authentication, although in all cases it is heavily encouraged. While cookies can be used to track internet behaviour, a much richer picture of the user activity can be collated when a user has been formally identified through a Google account login; for example, when using Chrome, their data will be processed, and form part of the "activity" tied to the customer's Google account. Activity can encompass anything from location, browsing and search history, YouTube videos watched, surveys answered, news preferences and ad personalisation. Options exist for deleting this data, but very few users are aware or know how to do this. Google will then use this composite profile of the users derived from the activity data to tailor advertising.

An additional element in this puzzle is provided by Single-Sign-On (SSO), also referred to as social login. SSO is a type of identity management technology that emerged following the popularity of services like Google and Facebook. When logging in to a website, users can opt to utilise their Google credentials to speed up the authentication process. OpenID Connect (OIDC) technology ensures that Google never exchanges passwords with a website, which decreases the risks of hacking personal credentials. Customers can therefore authenticate more quickly, but this comes at a cost: privacy.

Embedded in single sign-on, a protocol called OAuth2 is used to manage the types of actions that a service is authorised to run on a user's behalf. For example, someone could authorise an app accessed via Google to automatically send emails when a marketing coupon becomes available for an item of interest. This simple scenario can however be deceiving, because the details around which specific actions are being authorised are not always clear to the users and, as we saw earlier with browser cookies, the default action is to accept the terms without paying due attention. In summary, the convenience of quicker authentication increases the risk of unchecked authorisation.

With this technology in place, Google has yet more to gain. In fact, the profiling already being performed through tracking searches, location, preferences and browsing history is now enriched with data coming directly from third-party websites or apps that customers have access via SSO. This provides Google with an even more granular picture of the user, which, as we will see later, advertisers will find invaluable when bidding for ad space.

This report has focused on the technologies deployed by Google to collect data through traditional desktop web activity. However, Google is also leveraging the authentication platform to gather from an ever-growing array of devices across all aspects of our lives. These range from the Android operating system, which runs 70\% of the world's mobile phones, smart assistants in which Google's market share (25\%) trails Amazon's Alexa (70\%) and more recently through Google Wear and the acquisition of Fitbit delivering a significant presence in fitness tracking market.

\section{Digital Marketing}

When evaluating the advertising market, we first need to consider the different phases of a customers purchasing journey. The methods used to engage the potential customers necessarily evolve from early-stage, developing brand awareness, through to ensuring purchase consideration and finally into retargeting customer with the objective of finalising a sale. To deliver this capability internet advertising can be segregated into two principal formats, display ads and search ads, in both of which Google occupies a near-monopolistic position. In this report, we primarily focus on display ads, used to engage specific users while they are surfing the internet, based on their profile and not, necessarily, directly related to their current activity.

\subsection{Structure of the Digital Marketing industry}

Like any market, digital advertising has a group of 'sellers' and a group of ‘buyers’. In this case the ‘sellers’ are the publishers, that make the advertising real estate available, while the 'buyers' are the companies wanting to gain online visibility through purchasing the right to display their adverts in these spaces. The ads exposed can have different shapes and serve several purposes, including developing brand awareness, advertising promotional offers or a new product launch.

Initially, static web pages, where large media outlets directly sold advertising space to companies in a manner reflective of traditional advertising, were the predominant form of marketing. However, increasing internet speeds, new web languages such a JavaScript and the launch of digital marketing exchanges enabled dynamic web pages. These dynamic pages were able to specifically target adverts to individuals at the point of delivery based upon their profile. This technology, known as real time bidding (RTB), now accounts for 86\% of the display advertising space, and it is here that Google has established a dominant position.

\begin{figure}[ht]
  \centering
  \includegraphics[scale=0.45]{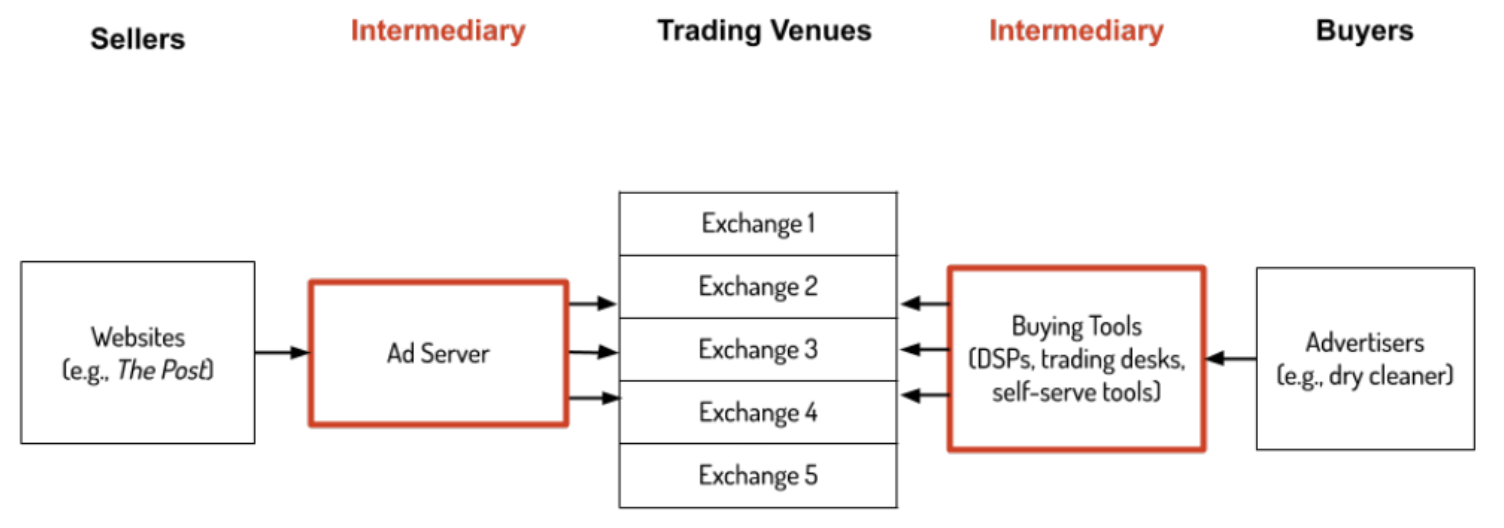}
  \caption{Marketing Industry Structure}
  \label{marketing_industry}
\end{figure}

Looking at the high-level landscape of the digital marketing industry as illustrated by \cite{srinivasan2020google} in Figure \ref{marketing_industry}, we see that this business model led to the creation of new actors, highlighted as "Intermediary". They can be seen as brokers, facilitating the exchange between Sellers and Buyers.

In this section of this report, we will explore each of the different actors and discuss how Google has managed to increase its footprint across the entire marketing value chain.

\subsection{Sell-side}\label{sec:sell-side}

The internet has grown exponentially since the 90’s with over 1.5 billion websites serving a wide variety of content. The universal access to the web has enabled small publishers to provide content via a website and gain exposure to a global audience. Advertisers quickly saw the potential of the web as an effective vehicle to deliver marketing campaigns to a wide audience. However, maintaining a website with accurate and engaging content can require significant resources for which advertising revenues are often the primary funding mechanism.

Initially, the nature of the contracts between publishers and advertisers was very much like classic bill- board advertising, with fixed (or guaranteed) contracts. But with the invention of the real time biding, publishers were able to sell ad space to advertisers in real time, based on the specific demographics of the user. The concept is rather simple: when a user visits the publisher’s website, the publisher sends information about the session to an Ad exchange. Depending on the information provided, the advertiser will assess the value and bid accordingly for the right to display an advert. The value an advertiser assigns to an ad space will be dependent on the perceived value and likelihood of the advert resulting in a conversion. To make this calculation as accurate as possible, it is crucial that the user profile provided is as rich as possible.

As we have explained previously, this information can be collected via various means, the main one being cookies. We can classify the cookies into two categories: the 1st party cookies placed on the user’s computer directly by the website visited and 3rd party cookies which are associated with other companies who indirectly populate the website content and therefore gain the right to place a cookie on the user’s computer via the publisher’s website. This concept of 3rd party cookies changed the landscape of the web, enabling user behaviour to be tracked across many websites, empowering the Ad network as known today.

As described earlier network effects mean that the more websites use a particular 3rd party cookie, the richer the user profile that can be created. In 2007 Google acquired DoubleClick, one of the biggest Ad networks at the time. Since then, Google has continued to grow its network by dispatching more and more cookies onto people’s computers. Google, along with Facebook, dominate the ad market, giving them an extensive understanding of the digital behaviour of millions of people. This deep understanding of people and the ability to target them increased the value of the service Google can offer. Google did not unlawfully gain their advantage in the Ad network space; they used their financial resources to both acquire capability and grow organically. We question however whether the dominant position they have now achieved is being used to stifle innovation from new entrants through maintaining a speed and information advantage others are unable to match.

One area of particular concern involves the practice of user-id hashing used to anonymise users for all but Google’s own platforms. This means that while Google can recognise users across multiple transactions, its competitors are not afforded the same privilege providing Google with a significant information advantage. Furthermore, this practise also introduces the possibility that other intermediaries can end up bidding against themselves.

\begin{figure}[ht]
  \centering
  \includegraphics[scale=0.75]{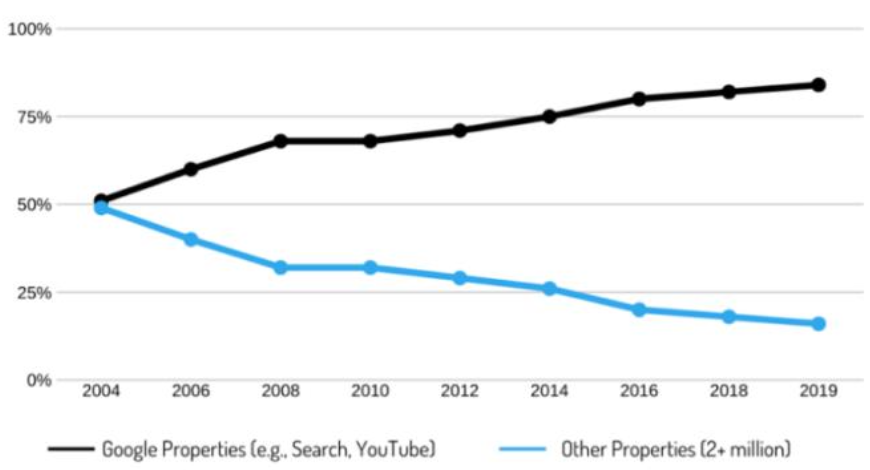}
  \caption{Distribution of allocated Ad space}
  \label{distribution_allocated_ad_space}
\end{figure}

To complete the picture of ad market dominance, it is important to consider the dominant role Google also has as a publisher. As described above, Google Maps delivers geographically targeted ads to millions of users while Google Search handles 92\% of the searches made online and constitutes the single largest digital ad space in the world. However, one other important component of the Google application suite not yet discussed is the popular video sharing platform YouTube. The website \cite{Oberlo_2021} states that YouTube has an active user-base of 2.3billion people and \cite{Oodle_2016} goes on to explain that traditional TV viewing is decreasing by 10\% a year, while YouTube is growing at a rate of 40\% with marketing spend following. Furthermore, Google withholds access to YouTube advertising space solely for its own ad-exchange, strengthening the tie for advertisers into its dominant marketing ecosystem. These factors combined have resulted in Google properties growing their share of digital marketing spend to a staggering 85\%, as reported by \cite{srinivasan2020google} and shown in Figure \ref{distribution_allocated_ad_space}.

\subsection{Ad Exchange}

The new era of digital marketing, fuelled by growing amounts of users’ data, enabled advertisers to make an educated decision regarding the value to assign to a specific impression. The profiling of the users is now a lot more precise, as it does not only focus on demographic attributes but also on behavioural events. The advertisers can leverage web history as a proxy to understand user behaviour. This had the mutual benefit of increasing the value of the advertising real estate while also achieving higher conversion rates for the advertisers. Furthermore, where the market was previously dominated by large organisations it is now far more accessible, allowing smaller publishers to have a place in the market.

The concept of an Ad exchange was proposed to ease the connection between publishers and advertisers. Similar to an electronic financial market buyers and sellers are performing their transactions programmatically in sub-seconds on one platform. The Ad Exchange orchestrates the auction process between publishers selling ad real-estate and the advertisers buying the right to display their ad. The Ad exchange runs an auction process (typically 1st price auction) to decide which ad will be shown on the publishers' website.

When Google acquired DoubleClick in 2007 it not only gained the ad selling intermediary business it also acquired a, yet to be launched, innovative digital exchange capability. With Googles strength and financial muscles AdX quickly grew to hold a 57\% of the market share and became the go-to exchange for publishers and advertisers. Google has been criticised many times for the lack of transparency in the way their system work and concerns raised about the fairness of the auction process. Allegations have also centred around the speed and information advantage that Google enjoys along with concerns regarding favouring their own publishing platforms.

In the financial market, electronic exchanges are independent entities guaranteeing quasi-fairness of the market and trust in the exchange. We can see that Google owning the principal exchange, even though it connects to others, raises a conflict of interest. Google plays a role as an Ad network selling web real-estate but also as a DSP, making them a big actor on both ends of the exchange. The digital ad exchange is a space where milliseconds are essential, as identified in the Martin Lewis book "Flash Boys" which highlighted the implications of latency in the trading process. Thus, the fact that Google owns the platform allows them to use its own data centres (collocate) to run the entire bidding process, giving them a significant advantage. According to a post by Google, up to 25\% of bids from non-Google buying tools fail due to latency issues compared to zero from their own collocated devices. The additional time at their disposal also enables them to perform a complete analysis and better determine the potential value of a user. Furthermore, if rival bids are excluded due to latency, effectively Google ads are able to acquire the ad space at a lower price.

Google Cloud does offer competitors the opportunity to collocate, which was taken up by OpenX in 2019. In promoting the deal, Google reiterated the competitive benefits. However, due to the lack of transparency, it remains challenging to quantify the benefit this really has. Unfortunately, due to the cost associated with running such services on Google Cloud, very few others have taken up this offer.

To promote competitiveness and derail Google's monopoly, a group of rival digital ad players developed an alternative technique called Header Bidding, offering the ability to connect to multiple ad exchanges, including Google's, concurrently. The JavaScript code running the auction process between the multiple exchanges was executed in the header of the web page of the publisher on the client machine.

The Header Bidding logic did, however, introduce slight delays in the loading of the page with a potential detrimental impact to the user experience. In response Google created a development framework called AMP, Accelerated Mobile Pages, which claimed to accelerate the delivery of rich content. In doing so, AMP imposed several restrictions on web page designs, one of which prevented the use of the JavaScript code required to perform header bidding. This policy, combined with modified criteria for appearing on the Google search engine, secured Google's dominant position. At this time Google also introduced Open Bidding, allowing the connection to other exchanges directly from their products, while avoiding the risk of their selling tools being bypassed.

\subsection{Buy-side}

The buy-side of the market corresponds to the advertisers: the entities who want to run a marketing campaign and are looking to gain visibility on the web. They buy the ad space that the publishers offer via the Ad exchange described in the section above. As we have seen earlier, the bidding process is a real time auction happening at lightning speed where the advertisers need to make a decision within a few milliseconds on the price they should bid for a given impression of their ad. To make this decision, the advertiser is equipped with the information provided by the sell-side about the session it is bidding on. This information, if linked with other data sources, can be invaluable to understand who the user is and whether or not it is a good target.

Performing this logic of understanding the session information and programmatically bidding according to predefined criteria is not trivial and very often outside of the core competencies of the advertisers. This is why advertisers often use an intermediary called a DSP (Demand Side Platform) to help them with their digital marketing campaign. The DSP offers a single interface to connect to multiple exchanges and provide the advertiser with the ability to select criteria to bid on. The DSP is then in charge of matching the criteria selected with the information available during the auction and deciding the best bid to make. If we leave aside the technical challenges, we see that the crux of the problem is really to understand the data provided during the auction in order to make an educated decision on the price to bid.

As cited earlier, Google has mastered the art of collecting data. Whether it is via its Ad network, the Chrome browser, Google maps or Google Authentication, Google is able to gather enough information to recreate a very accurate profile for a given user. Each user is assigned a unique user ID that is used by the DSP to retrieve the complete profile of the user. This is gold for any advertiser, as they now have the ability to be very specific in their targeting criteria. The marketing campaign used to be directed to an extensive demographic group of people, but with the richness of the profile that Google now offers, advertisers can market at the individual level.

As we could expect, this unique ID is only useful for Google's DSP (DV360), which makes this platform the most preferred one. In 2020, it was the number one DSP platform with 45\% of the advertisers spending on it, just ahead of The Trade Desk and Amazon Advertising \citep{AdExchanger_2020}. Today an ad space trading on Google exchange drops by 50\% when advertisers can't identify users associated with the ad space available. As \cite{srinivasan2020google} states, "a client of the double click ad server has to go through Google's exchange in order to efficiently buy more ads targeted to the same users because only Google tracks users by the same ID". It is also stated that "Google actively markets to advertisers and publishers as a synergy for using Google all the way through the trade."

In conclusion, even on the Buy-side, Google has imposed its dominance and has become the platform where advertisers should spend part of their marketing budget. The credit goes to a well-oiled machine, allowing Google to track a user throughout all the Google network (Google suite and Ad network), giving access to a very detailed profile of each individual. This in-depth information allows the advertisers to run a more targeted campaign, allowing them to spend their money more intelligently.

\section{Impact on business}

\begin{figure}[ht]
  \centering
  \includegraphics[scale=0.5]{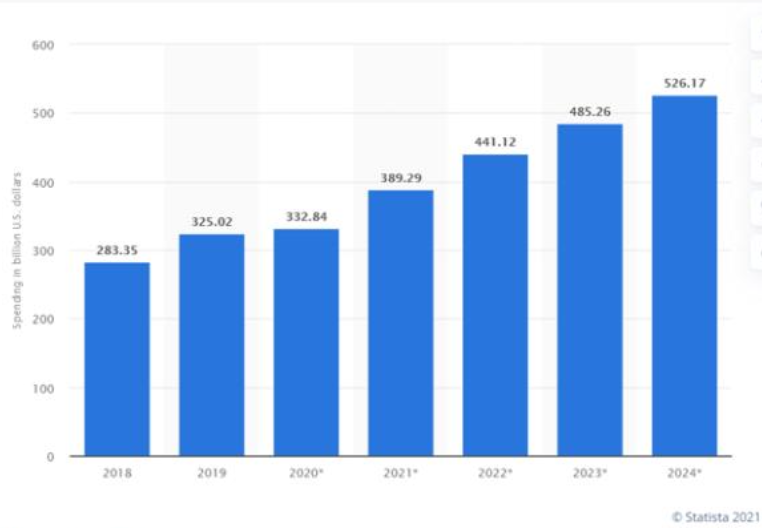}
  \caption{Growth of Global Digital Marketing Spend \citep{Statista_2021_digital_ads}}
  \label{growth_digital_marketing_spend}
\end{figure}

The speed and information advantage Google enjoy has allowed them to dominate the digital marketing industry and the broader monetisation of the internet. While the chart in Figure \ref{growth_digital_marketing_spend} from Statista illustrates the actual and projected growth of global advertising revenue, many sectors have seen a rapid decline as Google absorbs an ever-increasing proportion.

\subsection{The media industry}

\begin{figure}[ht]
  \centering
  \includegraphics[scale=0.5]{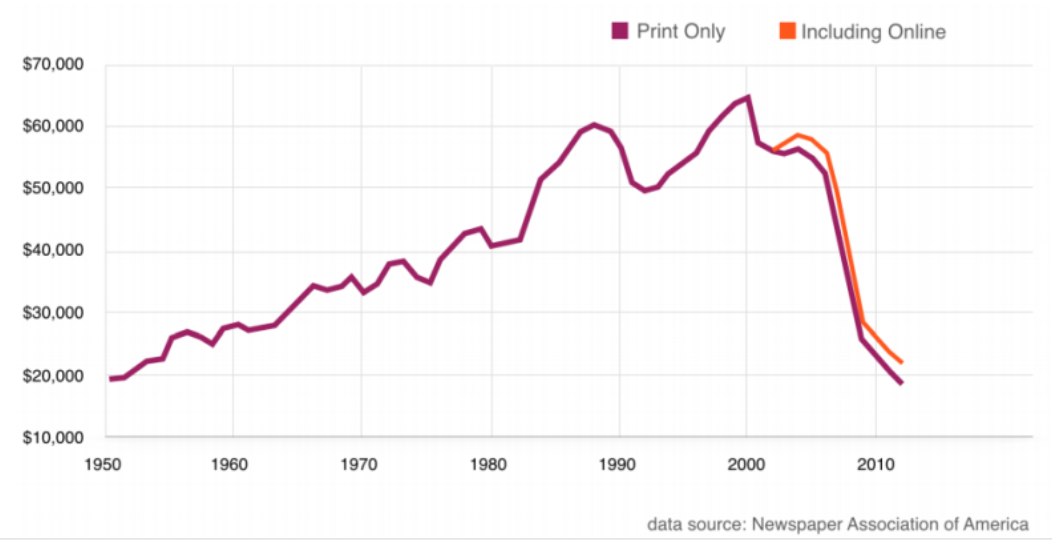}
  \caption{Decline Newspaper Ad Revenue (FIPP, 2019)}
  \label{newspaper_ad_revenue}
\end{figure}

Traditional media outlets have been particularly affected by Google's dominance. Their revenues have been shrinking (Figure \ref{newspaper_ad_revenue}) steadily, compromising their long-term viability and the content they create. As discussed in section \ref{sec:sell-side}, header bidding began to grow in popularity as an alternative to real time bidding around 2015. \cite{srinivasan2020google} suggest this immediately increased advertising revenues for many publishers by 50-70\% through increased competition and the removal of intermediaries; however, the actions taken by Google ensured this practice was quickly curtailed.

There is a significant risk that, without proper funding, high quality content on the internet will no longer be able to be produced. \cite{USAToday_2019} reports that “Between 1990 and 2016, there has been a 60\% reduction in the number of journalists being employed as people have increasingly sought to consume the news directly on Google's site, who didn't pay anything toward the cost of generating the content”.

Further supporting this the News Media Alliance \citep{TheGuardian_2019} reported that in “2018 Google made \$4.7 billion from news content alone which is the equivalent to that generated by every other news organisation in America combined”. The industry is trying to fight back and in 2019 managed to introduce the Journalism Competition and Preservation Act \citep{Congress_2019} which aims to allow “news companies to collectively negotiate with online content distributors regarding the terms on which the news companies' content may be distributed by online content distributors.” but the hard work is still to be done. The scale of the problem can be seen from the Guardians Media Group (GMG) financial report \citep{TheGuardian_2020}, which states:
\begin{itemize}
    \item Good growth in reader revenues offset by declines in advertising revenues and ongoing structural reductions in newsstand, both of which were exacerbated in the final month of the year due to the effects of coronavirus.
    \item Digital revenues of £125.9 million, up 0.5\% on the prior year (2019: £125.3 million). Digital revenues now make up 56\% of revenues.
\end{itemize}

\subsection{Parallels to the Microsoft case}

As reported in \cite{BusinessInsider_2020_antitrust} there are many similarities between the current lawsuits against Google and those faced by Microsoft 22 years ago. In both cases, the main accusation is that both companies are using their market power to create “a stranglehold on the industry and stifle competition”. In their defence both Bill gates, of Microsoft, and Sundar Pichai, from Google, state that they are the forefront of technology and fundamental to it is ongoing advancement.

At the time of the Microsoft case "network effects" were a relatively unheard-of term in an antitrust case and much of the case centred on the “insurmountable advantage this provided in crushing even superior products” provided by rival companies.

For Microsoft, the case centred around the dominance of its Windows operating system, citing agreements with IBM as potential collusion and the packaging of Internet Explorer for free as anti- competitive against the then principal alternative, Netscape. In the case of Google, it is its multifaceted dominance of the internet, agreements with Apple and Facebook and free offering of search, browser and other technologies to users that are drawing similar attention.

In the Microsoft case, the court initially ruled that the firm should be broken up into two separate companies; however, following an appeal, this judgement was overturned, imposing instead severe restrictions on the uncompetitive business practices that ultimately opened the door for the rise of Chrome as an alternative web browser. The question we seek to address here is: should similar actions be taken against Google to ensure the long-term growth and innovation of the internet?

\section{Summary and Proposal}

As we have identified throughout this report, Google has established a dominant position across the entire spectrum of the digital marketing industry through acquisition, innovation and, in some cases, questionable business practices. It now occupies a unique position operating as the principal publisher, demand and supply intermediary, as well as the most prominent ad exchange platform, resulting in significant conflicts of interest. The complexity and lack of transparency in the process have driven high transaction costs, reducing potential funding towards diverse and open journalism, entertainment, and other internet services, while also compromising user privacy.

Google have championed innovation and advancement of the internet through a range of free and trusted services, including browsers, search engines and maps. However, it is their stronghold on the marketing channels, which ultimately monetise most of the internet, creates a barrier to new entrants and thus furthering its dominance. Due to network effects, Google can extract more market value from acquisitions of rival technologies stifling longer term competition and innovation.

Furthermore, we argue that through this dominance, Google has achieved the role of unofficial regulator of the internet, such is the need to adhere to the requirements of Google's search engine. We also question whether exclusivity deals made with Apple and Mozilla, which locked more users into its search engine platform, where a legitimate means to ensuring it retained its near-monopolistic position.

As the widespread use of 3rd party cookies is due to end during 2022, the dominance of Google's own browser, maps and authentication platform will further serve to maintain their dominant position by developing a client-side alternative means of collecting the data supporting their information advantage in the marketing industry, unless new systems and regulations can be introduced.

\subsection{The role of the World Wide Web Consortium (W3C)}

Led by Tim Berners-Lee, founder of the internet, W3C is an international community that works to develop standards for the internet. W3C's mission is to "lead the World Wide Web to its full potential by developing protocols and guidelines that ensure the long-term growth of the Web." \citep{W3C_2020}

While the internet is seen by many as providing a wide range of free tools and services, these are primarily being financed by advertising revenues. We argue that to deliver the "long-term growth of the Web", it is fundamental that all market players can sustainably tap into the wealth generated by the internet. While the innovation and tools that Google has brought to the web have been crucial to its growth, the market dominance it now enjoys compromises the longer-term viability of the very content that people seek to consume.

While today W3C primarily focuses on establishing technical standards, we propose that it should take a broader remit to oversee the commercialisation of the web. The recent creation of the "Improving Web Advertising Business Group", with the remit of "improving the ecosystem and experience for users, advertisers, publishers, distributors, ad networks, agencies and others", is a step to address this requirement, but it still currently lacks the necessary authority to have a substantive impact.

The speed and informational advantage that Google has obtained in the marketing business through its dominance, network effects and economies of scale are exerting excessive control with a significant risk of distorting prices, creating barriers to innovation, and compromising user privacy, and it needs to be addressed.

We cite three specific areas of concern:
\begin{itemize}
    \item \textbf{Self-appointed unofficial regulatory powers:} The practice of excluding websites from its search engine that do not comply with its AMP framework, which in turn precludes the development of alternatives to its dominant ad server, is anti-competitive. This is exacerbated by the exclusivity deals made with Mozilla and Apple to ensure that Googles dominance of the search engine market is retained.
    \item \textbf{Transparency and fairness of the exchange:} Collocated exchanges form an anti-competitive mechanism as they provide a significant speed advantage. Furthermore, the lack of transparency of these transactions call into question the independence and fairness with which they operate, especially when Google is concurrently the dominant publisher, ad server and demand server.
    \item \textbf{Information sharing:} Hashing user identities for all but its own systems exacerbate the unfair information advantage, allowing Google to profile users to a level of precision that is inaccessible to its rivals.
\end{itemize}

\subsection{Blockchain as a potential solution}

A blockchain is a secure and verified digital record of transactions stored across a network of computers in a single list, known as a ledger. When a transaction is performed, it is transmitted to, and authenticated by, a network of computers. These transactions, known as blocks, are then chained together into a single list that forms a secure record. To change any record the amendment would have to be to all subsequent blocks in the chain, which provides an incredibly high level of security. This distributed ledger system ensures that there is no central system, which could be exploited by any single body and provides increased transparency of the transactions performed.

We propose that the creation and enforcement of a blockchain exchange would bring much-needed transparency, as it would not be tied directly to Google, and its inner workings would be publicly auditable by everybody involved.

\cite{10.1145/3359552} states that blockchain can be used to create "digital marketplaces allowing participants to make joint investments in shared infrastructure and digital public utilities without assigning market power to a platform operator, and are characterised by increased competition, lower barriers to entry, and a lower privacy risk".

They conclude their paper stating the blockchain is not a panacea to solve all market problems and outline the following four criteria for successful adoption.
\begin{enumerate}
    \item When last-mile problems are not severe, and digital verification can be implemented without assigning control to an intermediary.
    \item When the reduction in the cost of networking allows participants to allocate rents from a digital platform more efficiently between users, developers, and investors.
    \item When the combination of a reduction in both costs (verification and networking) allows for the definition of new types of digital assets and property rights.
    \item When there is a need for greater privacy and the ability for users to control when and how their data is accessed and used.
\end{enumerate}

We believe that all 4 of these criteria are applicable to the Digital Marketing industry and while this approach may introduce some inefficiencies it forms a key component in forming a fairer and more sustainable internet for all.

Through the codification of the rules under which the market operates no single party will be able to universally make changes that disadvantage other parties. The enhanced and shared visibility over the information generated by the network will remove the information advantage driven by network effects and in doing so will increase opportunities for innovation from new entrants.

In order to co-ordinate and maintain the network we envisage an expanded role to the W3C to both ensure that all internet and browser applications are compatible and supportive of the protocol. In particular, tools such as Google Search must not penalise websites adhering to the protocol and Google Chrome browser must support the population of the ledger in a consistent way.

\subsection{Basic Attention Token (BAT)}\label{sec:BAT}

At the heart of any currency there must be a unit of value. In traditional currencies this was originally underpinned by the gold standard. Most modern paper currencies are now derived from stability of the issuing government. \cite{Investopedia_2021_Bitcoins} define six conditions for a successful currency which are: scarcity, divisibility, utility, transportability, durability, and counterfeitability. The article goes on to explain how Bitcoin, the largest digital currency satisfies these requirements largely through the practice of Bitcoin mining.

In the case of digital marketing \citep{Brave_2021} introduces the concept of a measure of user attention called the BAT. Using the BAT as digital advertising token a decentralised ad exchange can then be created allowing advertisers, publishers, and users to connect directly. Launched in February 2021, Brave provides a dedicated open-source browser developed specifically to measures user attention. This method has the additional benefit of keeping the data on the users' device ensures privacy is maintained.

The flow of tokens from advertiser to publisher is illustrated in the diagram below, taken from the above paper. Interestingly in the article Brave also outlines future plans to allow users to redeem accrued tokens to unlock “premium” content from publishers’ websites, further unlocking future revenue generating opportunities.

\begin{figure}[ht]
  \centering
  \includegraphics[scale=0.75]{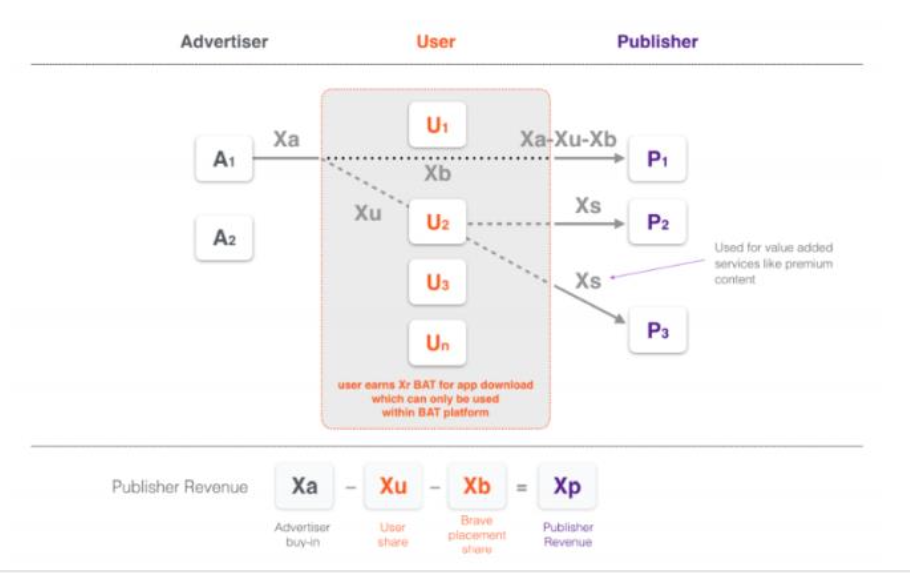}
  \caption{BAT Value Flow}
  \label{bat_value_flow}
\end{figure}

To date Brave reports an active user base of 25 million people, but to challenge Googles position far broader adoption would be required. In blockchain industries the bootstrapping phase, in which the primary concern is to secure a sufficient user base to deliver the required market value, is often cited as a critical hurdle in full scale adoption. It is here that W3C, with a more substantial presence, could accelerate the pace of adoption while also ensuring the appropriate distribution of resources to the wider internet ecosystem.

\section{Conclusion and further analysis}

Through this report we have consolidated the findings from multiple literature sources and have concluded that Google's practices have indeed driven significant innovation in the internet. However, the dominant position it now enjoys, particularly regarding the monetisation of the web has granted it an unfair advantage at the expense of rival publishers, service providers and user privacy. As an ever- increasing proportion of the wealth generated on the internet falls into the hands of a few, the very content people seek to consume is under threat. We have then recommended actions to address this issue, pointing to a larger regulatory role of W3C in support of the adoption of blockchain and new technologies such as the Basic Attention Token (BAT) as a measure of value.

Although only briefly addressed in this paper, it is worth reemphasising that Google has an even greater control in the mobile industry where, it not only controls many of the data gathering applications, but also controls the dominant operating system with a market share of over 70\%, according to data from \cite{Statista_2021_Mobile}. Through these mobile devices Google locks even more users into the authentication platform and collects an even richer data set as users, largely unaware, share personal information such as location and even health metrics on a regular basis. Unless action is taken now the goals of W3C to "lead the World Wide Web to its full potential by developing protocols and guidelines that ensure the long-term growth of the Web" are under significant threat.

\clearpage

\bibliographystyle{apalike}
\bibliography{digital_marketing}

\end{document}